\documentstyle[twocolumn,aps,prl,epsf]{revtex}% Use this for style revtex
\newcommand{\beq}{\begin{equation}}
\newcommand{\eeq}{\end{equation}}
\newcommand{\beqa}{\begin{eqnarray}}
\newcommand{\eeqa}{\end{eqnarray}}

\begin{document}

\def\rl{\rangle \langle}
\def\openone{\leavevmode\hbox{\small1\kern-3.8pt\normalsize1}}
\def\RR{{\rm I\kern-.2emR}}
\def\tr{{\rm tr}}
\def\ce{{\cal E}}
\def\cc{{\cal C}}
\def\ci{{\cal I}}
\def\cd{{\cal D}}
\def\cb{{\cal B}}
\def\cn{{\cal N}}
\def\ct{{\cal T}}
\def\cf{{\cal F}}
\def\ca{{\cal A}}
\def\cg{{\cal G}}
\def\cv{{\cal V}}
\def\cc{{\cal C}}
\def\rhon{\rho^{\otimes n}}
\def\on{^{\otimes n}}
\def\pn{^{(n)}}
\def\pnp{^{(n)'}}
\def\tcd{\tilde{\cal D}}
\def\tcn{\tilde{\cal N}}
\def\tct{\tilde{\cal T}}
\def\id{\frac{I}{d}}
\def\pthang{\frac{P}{\sqrt{d \tr P^2}}}
\def\psirq{\psi^{RQ}}
\def\rhorq{\rho^{RQ}}
\def\rhorqp{\rho^{RQ'}}
\def\ra{\rangle}
\def\la{\langle}
\def\z2{\{0,1\}}
%%%IEEE proof style
\def\QED{\mbox{\rule[0pt]{1.5ex}{1.5ex}}}
\def\proof{\noindent\hspace{2em}{\it Proof: }}
\def\endproof{\hspace*{\fill}~\QED\par\endtrivlist\unskip}

%%%new commands from john & barbara
\newcommand{\half}{\mbox{$\textstyle \frac{1}{2}$} }
\newcommand{\ket}[1]{| #1 \rangle}
\newcommand{\bra}[1]{\langle #1 |}
\newcommand{\proj}[1]{\ket{#1}\! \bra{#1}}
\newcommand{\outerp}[2]{\ket{#1}\! \bra{#2}}
\newcommand{\inner}[2]{ \langle #1 | #2 \rangle}
\newcommand{\melement}[2]{ \langle #1 | #2 | #1 \rangle}
\newcommand{\expect}[1]{\langle #1 \rangle}

%%%%%%%%%%%%%%%%%%%%%%%%%%%%%%%%%%%%%%%%%%%%%%%%%%%%%%%%%%%%%%%%%%%%%%%%%%%%%
\date{\today}
\title{A quantum circuit for OR}
\author{Howard Barnum$^{(1,3)}$, Herbert J. Bernstein$^{(1)}$, 
Lee Spector$^{(2)}$}
\address{
$^{(1)}$School of Natural Science and 
%Institute for Science and Interdisciplinary Studies (ISIS),
ISIS,
Hampshire College,
Amherst, MA 01002, USA \\
$^{(2)}$School of Cognitive Science, Hampshire College,
Amherst, MA 01002, USA \\
$^{(3)}$I.S.I. Foundation, Villa Gualino, Viale Settimio Severo 65, 
Torino 10133, ITALY \\
email: {\tt\{hbarnum, hbernstein, lspector\}@hampshire.edu}
}
\maketitle

%\centerline{Hampshire College and ISIS}
%\centerline{Amherst, MA 01002 USA}
%\centerline{hbarnum@hampshire.edu}
%\begin{flushright}
%\baselineskip=13pt
%\parbox{2.3in}{\baselineskip=13pt
%``This is an epigram''}\medskip\\
%---{\it Author of epigram}\\
%Source of epigram
%\end{flushright}

%\pagenumbering{arabic}
%\thispagestyle{empty}
%
%%%%%%%%%%%%begin text here
\begin{abstract}
We give the first quantum circuit 
for computing $f(0)$ OR $f(1)$ more reliably 
than is classically possible
with a single evaluation of the function.  
OR therefore joins
XOR (i.e. parity, $f(0) \oplus f(1)$) to give the full
set  of logical connectives (up to relabeling of inputs and outputs)
for which there is quantum speedup.  The XOR algorithm is of fundamental
importance in quantum computation;  our OR algorithm 
(found with the aid of genetic programming), may 
represent a new quantum computational effect, also useful as
a ``subroutine''.

% We present a quantum circuit evolved using genetic programming.  
% %With access, as a black box 
% %``oracle'', to a Boolean function of two bits, 
% From this array, we derive the first
% better-than-classical one-query bounded-error
% circuit for OR of one-bit black-box functions.
% The full evolved circuit calculates,
% with error probability lower than any possible classical one-query
% algorithm, the property defined by a depth-two 
% binary AND/OR tree with the four possible function 
% input values as leaves.  We analyze it as a kind of 
% recursive application of the OR circuit.
% Since the OR and AND/OR trees have fan-in $2$, these circuits may be 
% useful in investigating the uniform binary AND/OR tree in the large-$n$
% asymptotic regime, a problem whose classical query complexity is completely
% understood and which has 
\end{abstract}
\vspace{.1 in}
%, 89.70.+c}
\vspace{.2 in} \narrowtext
%\section{Introduction: genetic programming for quantum black-box computation}
All digital computing can be built out of Boolean functions.
Suppose 
a physical process
takes 
the two orthogonal inputs $\ket{0},\ket{1}$ to outputs 
$\ket{f(0)},\ket{f(1)}$, so that
$f$ is just a (classical) Boolean function from $\z2$ to $\z2$.  
This 
could be a quantum-coherent
computer subroutine, or the evolution of some other physical system we 
are interested in \cite{Preskill99a}.
If the process is a unitary quantum evolution, and we can
prepare a desired
state as input, then quantum computation lets one
find out more about the function $f$ than if
we are restricted to evolution of classical states.
For instance, we can find out its parity $f(0)$ XOR $f(1)$.
%(\cite{Cleve98a}, building on
%\cite{Deutsch85a}
%and \cite{Deutsch92a}). 
The XOR circuit in \cite{Deutsch85a} was the first concrete
demonstration of quantum computation's greater-than-classical power;
the exact version in \cite{Cleve98a} is of 
fundamental importance in its own right and
for its applications in more complex algorithms.

In this Letter, we complete the demonstration of quantum computation's
greater-than-classical power in this simple setting, by describing
circuits which
compute $f(0)$ OR $f(1)$ with one call to the subroutine $f$.  While 
these circuits, unlike the XOR circuit, may err, we show that their
performance is better than any possible classical circuit.  
This and XOR are the only quantum speedups in this simple
domain. (NOR, AND, NAND, and NOT-XOR may also be sped up
but they, and the algorithms that speed them up, are isomorphic to 
OR or XOR by simply
relabeling of inputs and/or outputs.)

The OR algorithm represents a new quantum computational effect,
which may have applications to more complex problems.
Indeed, our circuits
were derived as subroutines of a better-than-classical routine 
(evolved 
using genetic programming) to compute a more complex property, ``AND/OR,'' of 
Boolean black-box functions of two variables ($f: \z2^2 \rightarrow \z2$).
OR and AND/OR form part of an infinite family of properties
(uniform binary AND/OR trees) whose quantum
complexity is still imperfectly understood, but which have great theoretical
and practical importance in computing, since they have
applications in game tree evaluation, logic
programming, theorem-proving, and many other areas, and their classical
query complexity is completely understood.

The quantum complexity of Boolean properties of black-box functions
has been studied in \cite{Beals98a,Buhrman98a,Buhrman99b}.  
%Investigating other small oracle
%problems may help us understand the functioning and limits of quantum 
%computation.  
Here, we examine quantum gate
arrays for 
certain properties of black box functions of one and two qubits.  
%%Our work uses a new method for obtaining quantum gate arrays:  
%evolution via genetic 
%programming.
Given an unknown function $f$ which may be called on 
particular inputs or coherently on superpositions of them,
we wish to evaluate
a Boolean property $P$ of $f$.
We are interested in $p^e_{max}$,
the maximum over functions of the probability that
an algorithm evaluates $P(f)$ incorrectly, and $q_{max}$, the
maximum over functions of the expected number of function queries.
{\em Exact} algorithms have $p^e_{max}=0$;  
{\em Las Vegas} algorithms are correct whenever they answer
$0$ or $1$, but may also answer ``don't know'' with $p \le 1/2$.  
{\em Monte Carlo} algorithms
may err, but $p^e_{max} < 1/2$.  The error is {\em one-sided}
if there is a value $x$ ($0$ or $1$) such that $p_e=0$ for $f$
such that $P(f)=x$; otherwise it is {\em two-sided}.
By $n$ repetitions (and majority
voting, for Monte Carlo), the latter two types may be made to
have exponentially small (in $n$)
probability of not giving the correct
answer.
Below, we use a definition in which 
Las Vegas algorithms may have stochastic runtime, but
give correct answer with $p=1$;  
DFP (described below) is an example.
By running it repeatedly until an answer is obtained,
the first type of Las Vegas algorithm may be converted
into one of the second
type 
with expected running time greater by a constant factor.

%Using genetic programming (GP), we obtained better-than-classical
%gate arrays for the properties OR and for an
%AND/OR tree, which may also
%be relevant to uniform binary AND/OR trees in the large-$n$
%limit.  

%\section{Genetic programming of quantum algorithms}
Genetic programming (GP) \cite{Koza92a} evolves 
a population of programs (in our case,
sequences of quantum gates) which
are randomly mutated, recombined with each other, and  
preferentially selected for desired properties
by running (or simulating) the
programs on a sample of inputs.  
%Genetic programming searches over
%different structures for the gate array, with few constraints on the
%structures that may evolve.
The repertoire of gates used by our GP engines
(square brackets indicate qubit references, 
parentheses real parameters) was:
${\tt HADAMARD [q]}$, 
{\tt U-THETA[q](}$\theta${\tt)}$ := \cos\theta I + i \sin\theta \sigma_y$,
${\tt U2[q](\alpha, \theta, \phi, \psi}) := 
e^{i \alpha} 
e^{-i\phi \sigma_z} 
(\cos{\theta}I   -  i \sin{\theta} \sigma_y)
e^{-i\psi\sigma_z};~~{\tt CNOT[control, target]}$;
${\tt CPHASE[control,target](\alpha)}$
% \footnote{Note
%that this is a different form of {\tt CPHASE} than was used in our
%previous work \cite{Spector98a,Spector99a}.}
,
which multiplies each standard basis state by $e^{i\alpha}$
if it has $1$ in both control and target positions;
${\tt ORACLE [q_1,...q_n,q_{out}]}$, 
which adds (mod 2) $f(q_1,...,q_n)$ to $q_{out}$, retaining $q_1,...q_n$
unchanged; and 
{\tt MEASURE-0 [q]} and {\tt MEASURE-1 [q]}, which  
measure qubit $q$ in the standard basis.  The
{\tt MEASURE-x} gates terminate the computation, returning the value
$x$, if the measurement result $x$ is obtained;  if the result $\overline{x}$ 
is obtained,
the state is projected onto the subspace with 
$|\overline{x}\ra$ 
for that qubit, and computation proceeds.

Allowing termination
conditional on intermediate results, as {\tt MEASURE} gates
do, makes the number of queries stochastic.
For Monte Carlo algorithms, this yields at best a constant speedup
over algorithms with a definite number of queries.  Nevertheless, it
may yield more perspicuous algorithms, and 
the constant speedup may be needed for better-than-classical
performance, especially in the small-$n$ regime.  
%Earlier GP-derived results 
%were reported in \cite{Spector98a,Spector99a}.

%\section{The AND/OR problem}  
For functions of binary 
strings of length $d$, the AND/OR 
problem is to evaluate a binary tree, having 
AND at the root and $d$ 
layers of alternating
OR and AND as nodes, with a $d + 1$st layer of $n \equiv 2^d$
leaves consisting of 
the values of the black-box function ordered by their input string
(viewed as a binary integer).  
This and related problems have many applications, 
for example in game-tree evaluation, dynamic programming, 
waveform analysis, and theorem-proving.  
%For even depth, the tree 
%is a uniform binary tree of NORs;  the family of uniform binary NOR 
%trees has the same complexity. 
% for complexity 
% measures that do not involve one-sided error, since the odd-depth cases of
% the NOR tree take the same values as
% the corresponding AND/OR tree with negation of inputs.)
%For simplicity, we describe the best classical Las Vegas algorithm for
%for the 
%binary NOR tree.  
Saks and Wigderson showed that ``depth-first pruning'' (DFP) is
the best classical Las Vegas algorithm for AND/OR
\cite{Saks86a}.  DFP uses a routine {\tt eval(node)} which 
returns the value of the node if it is a leaf, 
and otherwise randomly chooses a 
daughter of the node and calls itself on the subtree rooted
at that daughter.  If this call returns a value for the subtree that
determines the value of the node (1 if it is an OR node,  0 if it
is an AND node), {\tt eval} returns the appropriate value;  
otherwise it calls itself on the other highest-level
subtree of the node, and 
returns the value of that subtree.  DFP itself just calls {\tt eval(root)}.

%We assume the nodes are indexed by the values of the
%variable {\tt node}, with the root indexed by $0$;
%$f({\tt node}),$ defined on leaves only, is the black box function.
% The
% routine also has access to the relation {\tt daughter} (and the function
% {\tt sibling} which can be defined in terms of {\tt daughter}) 
% which gives the tree structure.
% It uses the following recursive routine:

% \noindent
% \newline
% {\tt
% subroutine evaluate(node)
% \newline
% if leaf(node) then
% \newline
% \indent if f(node) = 1 return 0
% \newline
% \indent if f(node) = 0 return NOT(f(sibling(node)))
% \newline
% \indent endif
% \newline
% else 
% \newline
% \indent node' = a daughter of node, randomly chosen with p=1/2.
% \newline
% \indent call evaluate(node')
% \newline
% endif
% \newline
% }
%The algorithm simply calls {\tt evaluate} on the root.
%Saks and Wigderson \cite{Saks86a} 
%showed it was the
%optimal Las Vegas algorithm for any depth tree.      

Santha\cite{Santha91a} showed, for 
read-once Boolean functions (for which there
is a Boolean formula containing each variable at most once), 
that
no classical Monte Carlo algorithm 
with all error probabilities below $p$ 
can have expected queries $q < (1-xp)Q$, where $Q$ is the time taken by 
the optimal Las Vegas algorithm, and $x=1,2$ as the error is one- or
two-sided.  (It is not known whether a quantum analogue of this holds.)
This is just the trivial 
speedup obtained by flipping a biased coin
to decide whether to do the optimal Las Vegas algorithm or output a
random bit (two-sided) or a zero (one-sided).
%, choosing the bias to achieve the desired error probability
%.)   
Thus a $q$-query quantum algorithm would
have to have
$p^e_{max} < \frac{1}{x}(1 - \frac{q}{Q})$ to be better-than-classical.  
DFP has worst-case expected queries 3 for
depth-two AND/OR, so a one-query quantum algorithm would need
$p<1/3$ two-sided, $p<2/3$ one-sided to do better than classical.
% , while
% a two-query algorithm with $p<1/6$ two-sided or $p<1/3$ one-sided would
% also be of interest.
There is no one-query, zero-error quantum strategy
for calculating OR for a black box Boolean function of one bit
\cite{Jozsa91a,Beals98a}.
DFP has expected queries $3/2$,
so a one-query quantum algorithm with 
$p<1/6$ two-sided or $p<1/3$ one-sided would be better than classical.

%\section{Algorithms for depth one and two AND/OR}

Our OR circuit is shown in Fig. 2.
We use the
convention that 
the leftmost qubit in a ket or string of kets 
is qubit $0$, the next qubit $1$, and 
so on.
Before the {\tt MEASURE-1} gate on qubit $0$, the state is:
\beq\label{eq: state before measurement}
(1/2) ~[~ \ket{0} (\ket{f_0} + |f_1\ra) + \ket{1}(|f_0\ra - |f_1\ra) ~]\;.
\eeq
Thus the {\tt MEASURE-1} has $p=0$ for outcome $1$ if 
$f_0 =  f_1$ (even parity), due to destructive interference,
while if the parity is odd, it
has $p=1/2$ of correctly 
yielding $1$ ($|f_0\rangle$ and $\ket{f_1}$ are orthogonal, 
and do not interfere).
If the computation does not halt, the state becomes:
\beq
(1/2)\ket{0}(\ket{f_0} + \ket{f_1})\; .
\eeq
Its squared norm is the probability
that the {\tt MEASURE-1} gate yielded $0$.
First consider $\theta = 0$.
For the even parity cases, this term gives the correct answer, 
while for the odd parity cases, it is equally likely
to give either answer;  it contributes $1/4$ to $p_e$.  Thus 
$p^{00}_e=p_e^{11}=0$, and $p^{01}_e=p_e^{10}=1/4$.
The error is one-sided, so it is better than
classical ($p_e^{max} < 1/3$).  (If we had not halted the computation
when the measurement of qubit $0$ yielded $0$, 
and had measured qubit $1$ in the eigenbasis of $\sigma_x$
instead of the $\sigma_z$-eigenbasis 
used in the $\theta=0$ version of our 
algorithm, one sees from 
(\ref{eq: state before measurement}) that a value of $0$ for the final
measurement means the value of the first measurement gives the parity,
while a value $1$ for the final measurement means the value of the first
measurement is noise.  This is Deutsch's Las Vegas algorithm for
parity \cite{Deutsch85a}.) 
Our algorithm also outperforms attempts to use 2-alternative Grover 
search to evaluate OR; despite that method being asymptotically
optimal for OR of many inputs, it does not perform better than classically
in this instance.
%But that adaptation, which is also 
%the one-query version of the Deutsch-Jozsa algorithm for PARITY, but with 
%even parity interpreted as $0$, has $p^e_{11}=1$;  our algorithm 
%thus differs from it also.)
%With $X(\theta)$ before the final
%measurement we have:
%\beqa
Adding an $X(\theta)$ before the final measurement gives:
$
p_e^{odd} = 1/2 + \bigl( \frac{c - s}{2} \bigr)^2\;,
~p_e^{even} = s^2\;.
$
%\eeqa
%Setting these equal and taking square roots, we obtain the solutions:
Equating these gives a solution 
%$$
%c = \{-3s, s\} 
%$$
%The solution
%$$
$
c = \frac{-3}{\sqrt{10}}, s = \frac{1}{\sqrt{10}}\;
$
%$$
%is superior, and yields 
with $p_e= 0.1$ for all cases.
Since $p_e < 1/6$ two-sided with one query, this is also better than classical.

%This is better than the best classical one-query algorithm with
%two-sided error, which has 

% Which value of $\theta$ is more useful in a given context
% will of course depend on our cost function;  there may be some cases
% where the fact that the $\theta = 0$ yields zero error for $00$ and
% $11$ may make it superior, perhaps as a part of a routine for a 
% greater-depth AND/OR.

%\subsection{Optimality of the one-qubit array}
To see that the $\theta = 0$ array for OR minimizes
$p^e_{max}$ subject to the constraint
$p^e_{00}=p^e_{11}=0,$
consider the state 
%of the computer j
just 
before the black-box function is queried:
%Consider the pure state of a quantum computer (and its ancilla or 
%environment if necessary) running a putative algorithm
%for OR, just before the black-box function is queried.  This may
%be written:
\beqa
|\Psi \rangle  =  |\psi_{00}\ra |00\ra  +
|\psi_{01}\ra |01\ra  +
|\psi_{10}\ra |10\ra  +
|\psi_{11}\ra |11\ra \;.
\eeqa
The right-hand ket in each term is a state of the two qubits
on which the function will be called, and 
%states of the rest of the computer (``ancilla'').
%The $|\psi_{ij}\ra$ are possibly 
%subnormalized and/or non-orthogonal;  they 
%satisfy
%\label{normalization}
$\sum_{ij} \la \psi_{ij} | \psi_{ij} \ra = 1.$  
%\eeq
After the query, the state is:
\beqa
|\psi_{00}\ra |0f(0)\ra +
|\psi_{01}\ra |0\overline{f(0)}\ra +  
|\psi_{10}\ra |1f(1)\ra +   
|\psi_{11}\ra |1\overline{f(1)}\ra \;. \nonumber
\eeqa

The four functions of one bit give
states $|0\ra, \ket{1},\ket{2},\ket{3}$.
% \begin{center}
% \begin{tabular}{c|c|c} 
%  $f_0f_1$ & {OR}  & {State after $f$ called} \\ 
% \hline
% 00 & 0 &
% $~~~~~|\psi_{00}\ra |00\ra  +
% |\psi_{01}\ra |01\ra  +
% |\psi_{10}\ra |10\ra  +
% |\psi_{11}\ra |11\ra~~~~~ $ \\
% 01 & 1 &
% $~~~~~|\psi_{00}\ra |00\ra  +
% |\psi_{01}\ra |01\ra  +
% |\psi_{10}\ra |11\ra  +
% |\psi_{11}\ra |10\ra~~~~~ $ \\
% 10 & 1 &
% $~~~~~|\psi_{00}\ra |01\ra  +
% |\psi_{01}\ra |00\ra  +
% |\psi_{10}\ra |10\ra  +
% |\psi_{11}\ra |11\ra~~~~~ $ \\
% 11 & 1 &
% $~~~~~|\psi_{00}\ra |01\ra  +
% |\psi_{01}\ra |00\ra  +
% |\psi_{10}\ra |11\ra  +
% |\psi_{11}\ra |10\ra~~~~~ $
% \end{tabular}
% \end{center}
%
%Call the four states in the right hand column $|1\ra,|2\ra,|3\ra,|4\ra$.
%Then:
We have:
\beqa
\inner{0}{1}&=&\inner{\psi_{00}}{ \psi_{00}} +
\inner{\psi_{01}}{ \psi_{01}}  + 2 {\rm Re}\inner{\psi_{10}}{\psi_{11}}\;. \\
\inner{0}{2}&=&\inner{\psi_{11}}{ \psi_{11}} +
\inner{\psi_{10}}{ \psi_{10}}  + 2 {\rm Re}\inner{\psi_{00}}{\psi_{01}}\;. \\
\inner{0}{3}&=&2 {\rm Re}\inner{\psi_{00}}{\psi_{01}}
+ 2 {\rm Re}\inner{\psi_{10}}{\psi_{11}}\;.
\eeqa

%To be able to determine, with no error, the value of
%$OR$, 
%it must be able to perfectly distinguish $|1\ra$ from $|2\ra$,
%$|3\ra$, and $|4\ra$;  thus $|1\ra$ must be orthogonal to each of the
%others, 
%we need
%$\inner{1}{j} = 0 \; ; ~j=2,3,4,$ but this is impossible.
%These equations are incompatible with the requirement that
%some $|\psi_{ij}\ra$ be nonzero (cf. \ref{normalization}), and therefore
%no exact one-query routine exists.
Requiring $p^e_{11}=0$ and thus $\inner{0}{3} = 0$, 
there are optimal algorithms which
measure $\proj{0}$ and $I- \proj{0}$. 
For these,
$p_e^{max} = \max_{j=1,2,3}|\inner{0}{j}|^2$.
%Our algorithm amounts to setting $|\psi_{01}\rangle 
%= |\psi_{11}\rangle = 0$
%and $|\psi_{00}\rangle = |\psi_{10}\rangle = (1/\sqrt{2})|\psi\rangle$ for
%a normalized state $\ket{\psi}$ of an (optional!) 
%ancilla.  
%Defining $x_{ij} := \inner{\psi_{ij}}{\psi_{ij}}$ and 
%$\mu := {\rm Re~}\inner{\psi_{10}}{\psi_{11}}$, we have:
%\beqa
%\inner{1}{2} & = & x_{00} + x_{01} + \mu \nonumber \\
%\inner{1}{3} & = & x_{10} + x_{11} - \mu\;. 
%\eeqa
%Define $\Sigma := x_{00} + x_{01}$;
%since $\sum_{ij} x_{ij} = 1$, 
%the inner products are $\Sigma + 2\mu,~~1 - \Sigma - 2\mu.$
%If these are both positive, the minimum
%error probability occurs when both
%are equal to $1/2$, at $p_e = 1/4$.  
A simple calculation using the Schwarz inequality and
$\sum_{ij} \la \psi_{ij} | \psi_{ij} \ra = 1$ shows that 
$p^e_{max}$ is minimized 
where
$\inner{0}{2}=\inner{0}{3} = 1/2$, at 
$p^e_{max}=1/4$.   
% [I need to explore why this doesn't
% seem to force $\mu=0$, as I'd hoped (I haven tried to use the Schwarz
% inequality constraint on $\mu$, but it hasn't helped much so far although
% I may have made some error), 
% and explain the trivial variant of the algorithm that has
% $\mu = 0$ but uses both $x_{i0}$ and $x_{i1}$ (the latter just keeps
% track, in the ancilla, of whether we are working with $f$ or $f$ 
% complemented, and obviously we can later measure that and take it
% into account in interpreting the output, with no damage.)  Also, I need
% to deal with the possibility of one of the inner products being 
% negative.

%\subsection{A geometric view}
%A geometric point of view may yield some more insight.
% into how the depth-one
%version works. 
% Consider the four vectors $\ket{\psi_{00}}=|00\ra,\ket{\psi_{01}}
% =(1/2)~[~|0\ra (|0 \ra + |1\ra) + |1\ra (|0\ra - |1 \ra)~]~,
% \ket{\psi_{10}}=(1/2) ~[~ |0\ra(|0 \ra + |1\ra) - |1\ra (|0\ra - |1 \ra)~]~,
% \ket{\psi_{11}} = |01\ra$ which arise after the function call and the 
% Hadamard on $0$, when the function
% is
%$00$, $01$, etc.  
% These are:
% \newline
% \newline
% \begin{tabular}{|l|l|}
% \hline
% Function ($f_0f_1$)  & State $|\Psi_{f_0 f_1}\rangle$ \\
% \hline
% $00$ &    $$ \\
% $01$ & $(1/2)~[~|0\ra (|0 \ra + |1\ra) + |1\ra (|0\ra - |1 \ra)~]~$ \\
% $10$ & $(1/2) ~[~ |0\ra(|0 \ra + |1\ra) - |1\ra (|0\ra - |1 \ra)~]~$ \\
% $11$ & $|01\ra$
% \\
% \hline
% \end{tabular}
% \newline
%The geometry of these vectors is easily understood from their inner 
%products:  
For our algorithm 
$\inner{0}{3}=\inner{1}{2}=0,$ while the other inner products are 
$1/2$.
The states span a 
3-d real subspace of the 4-d complex space
of two qubits.  
% If we take two planes at 90 degrees to each other,
% intersecting in a line passing 
% through the origin,  $\ket{0}$ and $\ket{3}$ will lie in one
% (``even-parity'')
% plane,
% at $45$ degrees on either side of the other (``odd-parity'')
% plane, while $1$ and
% $2$ will lie in the odd-parity plane, at 45 degrees on either side of
%the even-parity plane.  
They lie, evenly spaced,
on a cone with apex at the origin and opening angle $\pi/2$.
%$\ket{1}$ and $\ket{4}$ are opposite each other, as are 
%$\ket{2}$ and $\ket{3}$.
The rest of the algorithm measures three 
orthogonal subspaces:
$\{\ket{10},\ket{11}\}$
(outcome $1$ for 
{\tt MEASURE-1} 0 ; algorithm returns 1), 
\{$|00\rangle$\} (outcome $0$ for final 
measurement on qubit 1; algorithm returns 0),
and 
\{$|01\rangle$\} (outcome $1$ for 
final measurement on qubit 1; algorithm returns 1).
The outcome $1$ for the algorithm corresponds to the
3-d
subspace perpendicular to $|00\ra$), 
while the outcome $0$ corresponds to \{$\ket{00}$\}.
When {\tt MEASURE-1} 0 yields $1$, qubit 1 
lies along
$|0\ra - |1\ra$, so within the 3-d space spanned by the 
possible computer states
the outcome ``1'' for {\tt MEASURE-1} 0 involves one dimension, 
and the result ``1'' for the algorithm two dimensions.
Our algorithm makes a further finegrained measurement within these
two dimensions, but 
we may avoid this
(which could affect
the results when the routine is called recursively)
by converting it into one
which measures a single qubit at the end.  
% We must
% distinguish
% $\{\ket{1}(\ket{0}-\ket{1}), \ket{01} \}$ from $\{\ket{00}\}$.  
% We include the unaccessed dimension 
% $\ket{1}(\ket{0}+\ket{1})$ 
% with $\ket{00}$.  
% {\tt CHADAMARD 0 1} transforms these subspaces into
% $\{\ket{11},\ket{01}\}, \{\ket{00},\ket{10}\}$, 
% distinguishable by 
% measuring qubit $1$.  
Such an algorithm
is: {\tt HADAMARD 0 ;
ORACLE 0 1 ;
HADAMARD 0 ;
CHADAMARD 0 1 ;
CONTROLLED X-THETA 0 1 {$\theta$} ;
MEASURE-0 1 ;
MEASURE-1 1}.

The OR algorithm was derived by restriction from a larger algorithm
found via genetic programming.
This algorithm computes the
depth-two
case (AND/OR$_2$) of evaluating:
$$
(f(00) \vee f(01)) \wedge (f(10) \vee f(11))\;.
$$  
With the number of oracle calls
fixed at one, and selection to minimize $p^e_{max}$, 
GP yielded 
the algorithm 
reported in 
\cite{Spector99a}, where more detail on the GP
engine used to evolve it may also be found.  

%\section{Depth-one algorithm} \label{sec: onebit}

Hand simplification and improvement yielded
%, involving changing angles to likely values, 
%combining some sequences of gates (chiefly 
%$U_\theta$'s $U_2$'s , and $H$'s) into single gates, and hand-optimizing
%the final rotation angle on qubit 2, yielded 
the gate array of Fig 1.
Here
{\tt X($\theta)$} 
%= U2($\phi=0, \theta, \psi=\pi/2,\alpha=\pi/2$)}, with
has the matrix ($c := \cos\theta, s := \sin\theta$, $\theta = .0749...$):
$$
\left[
\begin{array}{rr}
c & s \\
s & -c
%\cos{\theta} & \sin{\theta} \\
%\sin{\theta} & -\cos{\theta}
\end{array}
\right].
$$

%\begin{figure} \label{fig: andorsimplified}
%\caption{Hand-tuned algorithm for AND/OR}
%{\tt{~~} \newline

%U-THETA 0 {0.78540 [pi/4]} \newline
%HADAMARD 1 \newline
%ORACLE ORACLE-TT 0 1 2 \newline
%HADAMARD 1 \newline
%MEASURE-1 1 \newline
%HADAMARD 0 \newline
%MEASURE-0 0 \newline
%X-THETA 2 0.07500 \newline
%MEASURE-0 2 \newline
%MEASURE-1 2 \newline
%}
%\end{figure}

%\begin{figure}[h]
%\epsffile{fig1.eps}
%%FIGURE 1 REMOVED to fig1.tex, made into fig1.ps
%\caption{Hand-tuned version of evolved AND/OR; $\theta=0.74909$...}
%\label{fig: andorimproved}
%\end{figure}

This algorithm has error probabilities constant on 
orbits of the automorphism group, given in Table 
I.
%, which also reports
%error probabilities for another evolved algorithm 
%which emerged while we were simplifying the first.

%\begin{table}

\vskip 10pt
%{TABLE I: Error probabilities for hand-tuned simplified AND/OR algorithm}
\begin{center}
\begin{tabular}{|l|l|l|l|}
\hline
Orbit  &  $p_e$ & Orbit & $p_e$ \\ 
\hline
0 0 0 0  &  .00560 & 0 1 0 1  &  .28731 \\
0 0 0 1  &  .28731 & 1 1 0 1  &  .21269 \\
0 0 1 1  &  .21269 & 1 1 1 1  &  .00560  \\
\hline
\end{tabular}
\end{center}  
\vskip 4pt
Table I: Error probabilities (to 5 digits)
for hand-tuned simplified AND/OR algorithm
%\label{table: newandorsimplified}
%\end{table}
\vskip 10pt

The {\em automorphism group} of a property $P$ consists of 
those permutations $\sigma$ of its input variables
which leave its value
invariant for all assignments (all black-box functions).  
% That is,
% if $\sigma \in Aut(P)$,
% \beq
% \forall f,~~ P(f(0), ..., f(n)) = P(f(\sigma(0)), ... , f(\sigma(n)))\;.
% \eeq  
AND/OR$_2$ has four input variables
$f_0 \equiv f(00), f_1\equiv f(01), f_2, f_3$; its automorphism 
group is generated by
$(0 \leftrightarrow 1), (2 \leftrightarrow 3), 
(0 \leftrightarrow 2, 1 \leftrightarrow 3)$.  
%Application of 
%$\sigma \in Aut(P)$ to the input variables of
%a function yields a different
%function 
$Aut(P)$ acts on functions via 
$f^\sigma(x) := f(\sigma(x))$ 
%with $P(f^\sigma) = P(f)$.
% \eeq
% %omit the following if shortening
% Defining
% \beq
% \chi_\sigma: f \rightarrow f^\sigma, \;
% \eeq
% the mapping 
% \beq
% \sigma \rightarrow \chi_\sigma
% \eeq
% is a group homomorphism; thus the $\chi_\sigma$ are an
% action of the automorphism
% group on the set of black box functions.
%omit to here if shortnening
For AND/OR$_2$, the orbits
of this action may be labeled
by representative functions (written as strings 
%$f_{00} f_{01} f_{10} f_{11}
%\equiv 
$f_0 f_1 f_2 f_3$):
$0000_1$,
$0001_4$,
$0011_2$,
$0101_4$,
$1101_4$,
$1111_1$.
%\newline
%\newline
%\begin{tabular}{|l|l|}
%\hline
%Orbit & AND/OR value \\
%\hline
%$0000_1$ & 0 \\
%$0001_4$ & 0 \\ 
%$0011_2$ & 0 \\
%$0101_4$ & 1 \\
%$1101_4$ & 1 \\
%$1111_1$ & 1 \\  
%\hline
%\end{tabular}
%\newline
%\newline
%The strings specify representative functions; 
Subscripts indicate the 
number of functions in the orbit.
%The first three orbits have $P(f)=0$;  the last three have $P(f)=1$. 
Our algorithm also has this automorphism group:
the outcome probabilities for all its measurement gates
are constant on orbits of the group.
It is better than classical,
since $p_e^{max} < 1/3$.
%its worst- is below the Saks-Wigderson-Santha
%bound of $1/3$. 

%\section{Depth-two as nested depth-one}
The structure of the algorithm suggests examining its restriction
to qubit 1, since the states with $0$ versus
$1$ input in qubit $0$ are still orthogonal when qubit $1$
is measured, so don't 
interfere.
If we fix the input for qubit $0$ at $0$ and remove
qubit $0$ from the algorithm 
%(so that in particular, we 
%do not do the {\tt MEASURE-0} gate), 
we may consider the algorithm to 
use only two qubits and 
to apply to a new function $\tilde{f} := f(0 \cdot)$ 
defined by  
$\tilde{f}(0) = 
f(00)$ and $\tilde{f}(1) =
f(01)$.
Relabeling qubit $1$ as $0$, qubit $2$ as $1$ and
$\tilde{f}$ as $f$, we
get the circuit of Fig. 2 for computing 
$f(0)$ OR $f(1)$. 

%\ref{fig: onebitandor}.

%Fig. 2 removed to fig2.tex, fig2.ps

% \begin{figure} 
% \caption{An algorithm for OR}
%  {\tt{~~} \newline
%  HADAMARD 0 \newline
%  ORACLE ORACLE-TT 0 1 \newline
%  HADAMARD 0 \newline
%  MEASURE-1 0 \newline
%  X-THETA 1 {$\theta$} \newline
%  MEASURE-0 1 \newline
%  MEASURE-1 1 \newline
%  }
% \label{fig: onebitandor}
% \end{figure}

Similarly, fix a value $x$ for qubit $1$ in the depth-two algorithm 
and view it as an algorithm for AND operating on the one-bit function $f_x(y)$ 
given by $f(xy)$.  This is not identical to the array 
derived by applying De Morgan's law ({\tt NOT-(NOT-A OR NOT-B) $\equiv$ A
AND B}) to our OR algorithm, but its action on all black-box
functions is the same.
% [The following explanation should probably be omitted in any final
% version, or relegated to a footnote--agree?]
% The latter would start with a Hadamard instead
% of a $U(\pi/4)$, would add a QNOT to qubit 1 after the oracle call, and
% would {\tt MEASURE-1} instead of {\tt MEASURE-0} 
% (since it would be using the OR
% algorithm, but would then complement the classical values returned by the 
% measurement gates at the end of the algorithm, an option which is not
% included in the set of gates we currently use in the GP scheme.  Of course, 
% it is equivalent to move the final classical negation before the
% measurement,
% as a QNOT;  then the two QNOTs on qubit 1 would cancel.  The algorithm
% derived
% from the high-order input bit of the depth-two algorithm is just like this,
% except that it does {\tt MEASURE-0} instead of 
%{{\tt MEASURE-1}, NOT};  this is
% compensated
% for by the use of $U(\pi/4)$ instead of $H$ in preparing the input qubit,
% which just results in the states of qubit 1 relative to qubit 0 being 
% $0$ or $1$ being swapped at the stage just before the measurements;  hence
% it makes our interpretation of the state $|0\ra$ as the state which 
%  justifies returning ``0'' at that stage (versus the DeMorgan's law version's
% interpetation of $\ket{1}$ as the 
% state which justifies returning ``0'')
% correct.
%[end of section to be zapped in final version]
%The algorithm, in other words, is entirely equivalent to the one derived
%from
%applying  
%DeMorgan's law to the OR algorithm.  
So the depth-two algorithm can be loosely viewed as a
``recursive'' application of the 
depth-one algorithm (modified to give AND at the top level).
%(Extending the recursion to $d=3$ in the obvious way 
%gave $p_e^{max} > 1/2$.)
%This ``recursive'' structure has a peculiarly quantum feature, and also
%a peculiar feature arising from the presence of intermediate measurements.
%We may view the top level algorithm as applying the depth-one AND algorithm
%to the two OR nodes defined by different values of the high-order bit.  
Loosely speaking,  
it superposes values for qubit $0$ (as in the AND
algorithm), and
calls a ``function'' of qubit $1$;
which function depends on qubit $0$.  
This is not quite accurate for two reasons.
First, the lower-level algorithm returns a value to
the upper level 
only if the {\tt MEASURE-1 1}
does not halt the computation.  If {\tt MEASURE-1 1}
halts the computation, the superposition
of qubit-$0$ values provides a random value for that bit;
the lower level algorithm is effectively 
called on a randomly chosen marginal black-box function, and the
result returned as our algorithm's final output.
%(and performance suffered in 
%a modified version in which it was). 
Second, if the lower-level ``function 
call'' is not halted
%, there is an additional peculiarly quantum complication.
the ``function'' of $i$ called by the
top-level AND routine 
%is the array 
%for 
%calculating OR($f(i \cdot$));  
has 
an $f(i \cdot)$-dependent $4 \times 4$
unitary matrix which is not of the 
usual black-box form.  
The state after the function call and second {\tt HADAMARD 1}
is:
\beqa \label{red house}
(1/2\sqrt{2})~[~ \ket{00} (\ket{f_{00}} + \ket{f_{01}})
+ \ket{01} (\ket{f_{00}} - \ket{f_{01}}) \nonumber \\
- \ket{10} (\ket{f_{10}} + \ket{f_{11}})
- \ket{11} (\ket{f_{10}} - \ket{f_{11}})~]~\;.
\eeqa
This is a superposition, with coefficients $1/\sqrt{2}$ and
$-1/\sqrt{2}$, of states of the form (\ref{eq: state before measurement})
of qubits 1 and 2, with the qubit 0 
recording which marginal black box function, $f(0\cdot)$ or $
f(1\cdot)$, is involved.

% Here we can see that the input bits 0,1  simply label which of various
% possibilities occur in qubit 2.  Qubit 2 always contains the 
% sum of two kets each of which is $|0\ra$ or $|1\ra$ depending on the
% value of the function on a particular input.  Bit 0 determines the value of the 
% leftmost input bit to the function, while in each term both of the 
% values for the righmost input bit occur, but the relative sign of the 
% two resulting kets is given by the value of bit 1.  The {\tt MEASURE-1} gate
% thus yields $1$ with probability proportional to the sum of the squared
% norms of the two terms with a $|1\ra$ for qubit 1, and thus with the relative negative
% sign.  (The {\tt MEASURE-1} gate
% probabilities are invariant under the automorphism group, 
% since they depend only on the number of
% odd-parity subtrees, which is obviously 
% invariant under the generators of the 
% automorphism group.)
% This yields complete destructive interference
% when the two function values are equal;  otherwise it yields the term
% \beqa
% \frac{1}{2 \sqrt{2}} (-1)^x (-1)^{f(x0)} (|0\ra - |1\ra)  
% \eeqa
% where $x$ is the value of qubit 0 (the overall sign is 
% unimportant here but 
% it is relevant if this routine is used as part of a larger gate array).
% The latter sort of terms have squared norm $1/4$, and thus 
When 
$f(00) \ne f(01)$ or $f(10) \ne f(11)$, but not both;
(orbits 
$1101_4$ and $0001_4$), the
{\tt MEASURE-1} has $p=1/4$ of halting the computation with 
result $1$, an error if the orbit is $0001$.   
If both subtrees have odd parity ($0101_4$
only), $p=1/2$ of halting and (correctly) yielding 1, 
and if neither does 
($0000$, $1111$, $0011_2$), $p=0$.  
% That the probabilities $0$ or $1/4$
% of obtaining ``1'', from
% the two values of qubit $0$,
% simply add illustrates that this is effectively the low-order
%OR algorithm acting on a randomly chosen marginal black box function.
This measurement contributes $1/4$ to $p^e_{0001}$.

The state after the {\tt MEASURE-1} gate yields $0$ is
given by the terms with $\ket{00}$ and $\ket{01}$ in 
(\ref{red house}).
%:
%\beqa
%(1/2\sqrt{2})~[~
%\ket{00} (\ket{f_{00}} + \ket{f_{01}})
%- \ket{10} (\ket{f_{10}} + \ket{f_{11}}) ~]~\;.
%\eeqa
%Its squared norm is the probability
%that the {\tt MEASURE-1} gate yielded $0$.  
Hadamarding qubit 0 yields:
\beqa \label{power of soul}
(1/4) & ~[~&  |00\ra \bigl(\ket{f_{00}} + \ket{f_{01}} - |f_{10}\ra - |f_{11}\ra
\bigr) \nonumber \\
& + &
|10\ra \bigl(\ket{f_{00}} + \ket{f_{01}} + |f_{10}\ra + |f_{11}\ra \bigr)~]~
\;.
\eeqa
The {\tt MEASURE-0} gate will terminate the computation with the result $0$
with probability given by the squared norm of the first term in this state.
% $0000$, $1111$, and $0101$  give $p=0$ due to 
% complete destructive interference. 
% %For the
% %two palindromic representatives of $0101_4$ this involves 
% %cancellation between
% %terms differing in the low-order bit, despite the fact that we are
% %now in the ``higher-level'' part of the algorithm.   
% $1101$ and $0001$ leave two orthogonal terms after two 
% have cancelled, for squared norm
% $1/8$.  
% %(Here, the only interference is between different 
% %values of the high-order bit, as would occur in using the AND
% %array on a standard black box.)
% This is correct for $0001$, but erroneous
% for $1101$.  $0011$ has no cancellations, so two copies 
% of each of two orthogonal terms are left, contributing
% $1/2$ to the probability of the correct answer $0$.  
%%(The ``constructive 
%%interference'' here is between terms with different values for the 
%%low-order input bit.)
If measurement yields $1$, the final state
is given by the second term in (\ref{power of soul}).  
Then the probability of the final measurement on qubit $2$     
giving $1 (0)$ is 
\beqa
(1/16) (n_{1(0)}(f))^2\;,
\eeqa
where $n_x(f)$ is the number of inputs on which $f$ takes
the value $x$.
%\eeqa
% %Here we see constructive interference between different 
% %values of the low-order bit {\em and} the high-order
% %bit.
% %This is obviously invariant under the automorphism group
% %(since it is invariant under any permutation).
% This gives $p=1$ for the final measurement returning the 
% correct answer for the functions $0000$ and $1111$.  Contributions to 
% erroneous outcomes are:  1/16 for $0001$ to yield $1$, $1/16$ for
% $1101$ to yield $0$, $1/4$ for $0101$ to yield $0$, and 
% $1/4$ for $0011$ to yield $1$.
The total error probabilities (with their sources) are:
\vskip 10pt
\begin{center}
\begin{tabular}{|l|lr|}
\hline
Orbit  &  Error probability & \\ 
\hline
0 0 0 0  &  & 0  \\
\hline
0 0 0 1  &  1/4 ({\tt M-1 0}) + 1/16 ({\tt M-1 2}) = & 5/16  \\
\hline
0 0 1 1  &  1/4 ({\tt M-1 2}) & 1/4 \\
\hline
1 1 0 1  &  1/8 ({\tt M-0 1}) + 
1/16 ({\tt M-0 2}) = & 3/16  \\ 
\hline
0 1 0 1  &  1/4 ({\tt M-0 2}) & 1/4  \\
\hline
1 1 1 1  &  & 0 \\
\hline
\end{tabular}
\end{center}
\vskip 10pt

This is better than classical even without the final $X(\theta)$.
%Thus $p_e^{max}=5/16$, for the orbit
%$0001$ to erroneously give $1$.  Even without the final 
%then, this is better than classical.  
%We now elucidate
%the effect of that  
%rotation on qubit $2$ before the final measurement.  
For
$0001$, $1101$, $0101$ the final measurement may give 
either result.  We may decrease $p^e_{max}$ 
(here $p^e_{0001}$) by  
increasing the likelihood that the final measurement yields $0$,
at the cost of increasing $p^e_{0101}$ and $p^e_{1101}$ for which
$P(f)=1$, and also possibly $p^e_{0000}$ and
$p^e_{1111}$. 
%the other orbits ($0000$ and $1111$) which can reach the final
%measurement.
%on $0101$ and $1101$
%for which the correct answer is $1$, and also, possibly, of inducing
%errors in the 
%other cases which have a nonzero chance of reaching the
%final measurement ($0000$ and $1111$).  
%When another error probability reaches 
%equality with $p^e_{0001}$, there will be no further improvement
%possible, at least by varying a single parameter..
Thus, consider adding, before the measurement, an $X(\theta)$ gate 
% $$
% \left[
% \begin{array}{rr}
% c & s \\
% s & -c
% \end{array}
% \right]
% $$
on 
qubit 2.
%With $c := \cos{\theta}$ and $s := \sin{\theta}$,
% For the $0001$ case, this effects:
% \beqa
% \frac{3}{4}|0\ra + \frac{1}{4}|1\ra \rightarrow
% \frac{1}{4}~[~(3c + s) |0\ra + (3s - c) |1\ra\;
% \eeqa
% for the $0011$ case, it effects:
% \beq
% \frac{1}{2}(|0\ra + |1\ra) \rightarrow \frac{1}{2}((c+s)|0\ra + (c-s)|1\ra)\;. 
% \eeq
We obtain
%\beqa
$p_{0001} = \left(\frac{3s - c}{4}\right)^2  + \frac{1}{4}$, 
$p_{0101} = \left( \frac{c + s}{2}\right)^2\;.$
%\eeqa
Equating these gives
%, one obtains the quadratic equation 
%$260 \eta^2 -180 \eta +1 = 0$ in $\eta :=   s^2$.
%The solution 
$s^2 = (9 - 14\sqrt{2/5})/26,$ so  
$\theta^* = 0.074909...$ at $p_e=p_{0001}=p_{0101} ~= 0.287315...$.
%(The other errors remain lower.)
% Moreover, while this increases the errors on $0000$ and $1111$ to
% $s^2$, and also increases the error on $1101$, these are still not
% maximal.  

%%\section{Conclusion}
With the help of genetic programming, we found better-than-classical
quantum gate arrays for
the depth one and two cases of the family of properties of black-box
functions given by alternating binary AND/OR trees.  These
circuits  
could constitute small-$n$ instances of a scalable Monte Carlo algorithm for
this family of  properties.
% Another interesting question related to this is whether a result analogous
% to Santha's classical result, that for read-once Boolean properties
% of black-box functions the only speedup Monte Carlo can achieve over
% Las Vegas is the trivial linear speedup, holds for quantum
% algorithms for some class of of Boolean properties as well.
They are also small-$n$ instances of the 
bounded-depth AND/OR tree problems whose 
complexity was characterized in 
\cite{Buhrman98a,Buhrman99b}.
The Grover-based algorithms therein achieve marked 
speedups over classical means
in the large-$n$ regime, as the fan-in of the nodes in the
bounded-depth trees grows; their $n=2$ and $n=4$ instances 
are not superclassical.  In contrast, our arrays give
speedups for fan-in $2$.  Since the uniform
binary tree problem has fan-in $2$ for all instances, 
this suggests that aspects of our 
gate arrays may prove useful in addressing this problem
even for large $n$.
If so, this would be another new quantum computational effect related
to the OR circuit.  As it stands, OR joins XOR to complete the
set of logical properties of one-bit
black-box subroutines which can be quantum-mechanically 
computed, with one subroutine
call,
more reliably than is classically possible.

%\appendix
%\section{No deterministic one-query quantum algorithm exists for OR on
%one-bit functions}
%\label{Appendix A}
\begin{acknowledgments}
%\section*{Acknowledgments}
Supported in part by the John D. and Catherine T. MacArthur Foundation's
MacArthur Chair program at Hampshire College, by NSF grant 
\#PHY-9722614, and by a grant from the ISI Foundation, 
Turin, Italy, and the Elsag-Bailey
corporation. 
% Some work reported here was performed at the ISI's 1998 Research
% Conference on Quantum Computation, supported by ISI and Elsag-Bailey.  
We thank H. Burhman, R. Cleve, M. Mosca, and
R. de Wolf for discussions.
% , and 
% Ronald de Wolf for suggesting the AND/OR tree problem and 
% providing references on it.
\end{acknowledgments}

%%%%%%%%%%%%%%%%%%%%%%%%%%%%%%%%%%%%%%%%%%%%%%%%%%%%%%%%%%%%%%%%%%%%%%%%%%%

%%%%%%%%%%%%%%%%%%%%%%%%%%%%%%%%%%%%%%%%%%%%%%%%%%%%%%%%%%%%%%%%%%%%%%%%%%%%%
%comment out when biblio is ready:
%\bibliographystyle{IEEE}
%\bibliographystyle{prsty}
%\bibliography{bib}

\begin{thebibliography}{10}

\bibitem{Preskill99a}
A.~M. Childs, J. Preskill, and J. Renes,   (1999), {L}os {A}lamos ArXiV
  Preprint Archive quant-ph/9904021.

\bibitem{Deutsch85a}
D. Deutsch, Proc R Soc London A {\bf 400},  97  (1985).

\bibitem{Cleve98a}
R. Cleve, A. Ekert, C. Macchiavello, and M. Mosca, Proc R Soc Lon A {\bf 454},
  339  (1998).

\bibitem{Beals98a}
R. Beals, H. Buhrman, R. Cleve, M. Mosca, and R. de~Wolf, FOCS '98  352
  (1998).

\bibitem{Buhrman98a}
H. Buhrman, R. Cleve, and A. Widgerson, Proceedings of the 30th Annual {ACM}
  Symposium on the Theory of Computing (STOC)  63  (1998).

\bibitem{Buhrman99b}
H. Buhrman, R. Cleve, R. de~Wolf, and C. Zalka,   (1999), {L}os {A}lamos ArXiV
  Preprint Archive cs.CC 9904019.

\bibitem{Koza92a}
J.~R. Koza, {\em Genetic Progamming: On the Programming of Computers by Means
  of Natural Selection} (The MIT Press, Cambridge, MA, 1992).

\bibitem{Saks86a}
M. Saks and A. Wigderson, Proceedings of the 27th {IEEE} Symposium on the
  Foundations of Computer Science ({FOCS})  29  (1986).

\bibitem{Santha91a}
M. Santha, Proceedings of the 6th {IEEE} Structure in Complexity Theory  180
  (1991).

\bibitem{Jozsa91a}
R. Jozsa, Proc R Soc London A {\bf 435},  563  (1991).

\bibitem{Spector99a}
L. Spector, H. Barnum, H. Bernstein, and N. Swamy,  in {\em Advances in Genetic
  Programming} (MIT Press, Cambridge, MA, to appear in 1999), Vol.~III.

\end{thebibliography}

%put biblio here when ready (filename.bbl):

\end{document}